# Rotating detonation combustion with *in-situ* evaporating bi-disperse *n*-heptane sprays

Shan Jin[1,2], Huangwei Zhang[2]*, Ningbo Zhao[1], and Hongtao Zheng[1]

1. College of Power and Energy Engineering, Harbin Engineering University, Harbin 150001, China
2. Department of Mechanical Engineering, National University of Singapore, 9 Engineering Drive 1, Singapore 117576, Republic of Singapore

## Abstract

Eulerian-Lagrangian simulations are conducted for two-dimensional Rotating Detonative Combustion (RDC) fueled by bi-disperse *n*-heptane sprays without any fuel pre-vaporization. Parametric studies are performed to study the influences of droplet diameter and droplet distribution on detonation wave are studied. The extinction process of the detonation wave is also been analyzed. It is found that small *n*-heptane droplets (e.g., 2 μm) are completely vaporized in the fuel refilling area. Increasing the droplet diameter causes the droplet to fail to evaporate completely within the fuel refilling area and exist after the detonation wave. A reflected shock can be obseverd after the detonation wave. When the droplet diameter is larger than 10 μm, the higher pressure after the detonation wave leads to the reactants cannot be sprayed into the combustor, eventually leading to extinction of the detonation wave. In bi-disperse *n*-heptane sprays, presence of droplets with small diameter maintains the stable propagation of the detonation wave. The average equivalence ratio in the fuel refilling area is lower than total equivalence ratio (up to 0.66 only), and the average equivalence ratio decreases with the increase of droplet diameter in bi-disperse *n*-heptane sprays. The increase in droplet diameter decreases the detonated fuel fraction and the speed of detonation wave. The speeds of detonation wave in bi-disperse n-heptane sprays are 3−9% lower than the respective gaseous cases. Moreover, propulsion performance of RDC such as thrust and specific impulse decreases with increasing droplet diameter.

*Corresponding author. E-mail: huangwei.zhang@nus.edu.sg. Tel: +65 6516 2557.

# 1. Introduction

Rotating Detonation Engine (RDE) is deemed one of the most promising pressure-gain combustion technologies due to the high thermodynamic cycle efficiency [1,2]. In previous studies, gaseous fuels are mainly tested, including hydrogen and simple hydrocarbons [1–5]. However, liquid fuels typically have a higher energy density and are more convenient to be stored and transported. Utilization of liquid fuels is of utmost importance to commercialize rotating detonation technology to engineering practice.

The first liquid fuel RDE test was carried out in 1960s [6], and in recent years a lot of research progress has been available for liquid fuel RDEs. For instance, Bykovskii et al. [5,7,8] used liquid kerosene sprays and oxygen-enriched air in experimental research on two-phase rotating detonation. The combustor diameter of outer wall in their test was 306 mm. They found that the addition of hydrogen to the mixture can reduce the critical diameter of the combustor. After that, they increased the diameter to 503 mm [7,8], and found that the Rotating Detonation Wave (RDW) cannot continuously propagate without hydrogen or syngas added.

In addition, Kindracki [9] investigated kerosene atomization characteristics under different nitrogen velocities and fuel injection patterns in a rotating detonation combustor. They measured the droplet diameters and found that when the sprayed droplet diameters are 20−40 μm, the droplets can quickly evaporate in the combustor. Subsequently, they used kerosene with hydrogen addition to conduct a two-phase rotating detonation experiment [10]. They successfully obtained a rotating detonation wave that propagates stably, with a velocity deficit of the detonation wave propagation being 20%−25%. More recently, Wolański and his co-workers [11] partially mixed the preheated liquid Jet-A and hot air, leading to a composition higher than the rich flammability limit. With that, they achieved a rotating detonation without hydrogen addition. They also found that heat losses, momentum losses and pre-combustion of the fuel are the main reasons for the speed deficit (up to 35%).

To obtain more detailed structures of rotating detonations in liquid fuel sprays, RDE modelers also carried out a series of numerical studies. For instance, Sun and Ma [12] investigated the effects of air total temperature and fuel inlet spacing on the two-phase RDW in liquid octane and air. They found that increasing the fuel inlet spacing decreases the wave speed. Moreover, Hayashi et al. [13] investigated the effects of JP-10 droplet diameter and pre-evaporation on two-phase rotating detonation waves. They found that there are liquid droplets along the contact surface between the fresh and burned gas. Ren and Zheng [14] studied the limit of kerosene/air two-phase detonation stability as a function of total pressure and total temperature considering the operating conditions of the ramjet detonation engine. They found that stable rotating detonation is achieved in a limited range of total pressure and increased total temperature is conducive to RDW stability.

Moreover, Meng et al. [15] used *n*-heptane/air as the reactants to systematically study the influences of initial droplet diameter (5-50 μm) and the *n*-heptane pre-evaporation degree on detonation characteristics (e.g., detonation velocity, detonated fuel fraction, droplet evaporation height). Meng et al. [16] also investigated the rotating detonation combustion with partially pre-vaporised *n*-heptane spray without hydrogen addition. They analyzed the detailed RDE flow field and droplet distribution inside the fuel refilling zone and found that a layer with high vapor concentration exists between the droplet-laden area and deflagration surface. Besides, Zhao and Zhang [17] investigated the influences of droplet diameter and equivalence ratio on rotating detonations. The propagation speed increases as the total equivalence ratio increases for the same droplet diameter. Furthermore, they observed that when the droplet diameter is less than 5 μm, the thrust force from pressure gain and kinetic energy decreases significantly with the droplet diameter. However, for initial droplet diameter $d_0 > 5$ μm, the thrust force from the kinetic energy first increases and then decreases with the droplet diameter, while the thrust force from pressure gain is shown to have limited change.

In this work, the effects of initial droplet diameter on wave speed, detonated fuel fraction and specific impulse in two-phase rotating detonation combustor will be further studied with Eulerian－Lagrangian method. In contrast to abovementioned studies, this paper focuses on the combustion process of liquid fuel droplets without pre-evaporation in the combustor (only *in-situ* gasification) and droplet bidispersity effects. Two-dimensional flatten domain is used to mimic the practical rotating detonation combustor, and liquid *n*-heptane and air are selected as the reactants. The rest of the manuscript is structured as below. In Section 2 the computational method and the physical model are introduced. Results are presented in Section 3 and conclusions are made in Section 4.

## 2. Mathematical and physical models

### 2.1 *Governing equation*

The Eulerian–Lagrangian method is used to investigate the two-phase rotating detonation combustion in this work. The gas phase is described with the Eulerian method, whilst the sprayed liquid fuel droplets are tracked by the Lagrangian method. For the gas phase, the governing equations for unsteady compressible multi-species reacting flows read [18]

$$\frac{\partial \rho}{\partial t} + \nabla \cdot [\rho \mathbf{u}] = S_m, \tag{1}$$

$$\frac{\partial (\rho \mathbf{u})}{\partial t} + \nabla \cdot [\mathbf{u}(\rho \mathbf{u})] + \nabla p + \nabla \cdot \mathbf{T} = \mathbf{S_F}, \tag{2}$$

$$\frac{\partial (\rho E)}{\partial t} + \nabla \cdot [\mathbf{u}(\rho E)] + \nabla \cdot [\mathbf{u}p] + \nabla \cdot [\mathbf{T} \cdot \mathbf{u}] + \nabla \cdot \mathbf{q} = \dot{\omega}_T + S_e, \tag{3}$$

$$\frac{\partial (\rho Y_m)}{\partial t} + \nabla \cdot [\mathbf{u}(\rho Y_m)] + \nabla \cdot \mathbf{s_m} = \dot{\omega}_m + S_{Y_m}, (m = 1, \dots M-1), \tag{4}$$

$$p = \rho RT. \tag{5}$$

Here *t* is time and $\nabla \cdot (\cdot)$ is divergence operator. $\rho$ is the gas density, **u** is the gas velocity vector, *T* is the gas temperature, and $p$ is the pressure. $Y_m$ is the mass fraction of $m$-th species, and $M$ is the total species number. $E$ is the total non-chemical energy, i.e., $E \equiv e_s + |\mathbf{u}|^2/2$. $e_s = h_s - p/\rho$ is the sensible internal energy and $h_s$ is sensible enthalpy. $R$ in Eq. (5) is the specific gas constant and is calculated from $R = R_u \sum_{m=1}^{M} Y_m MW_m^{-1}$. $MW_m$ is the molar weight of $m$-th species and $R_u = 8.314$ J/(mol·K) is the universal gas constant. Particle-source-in-cell (PSI-CELL) approach is used [16] and the source terms in Eqs. (1) – (4), i.e., $S_m$, $\mathbf{S}_F$, $S_e$ and $S_{Y_m}$, account for the exchanges of mass, momentum, energy, and species, respectively.

In Eq. (4), $\mathbf{s_m} = -D_m \nabla(\rho Y_m)$ is the species mass flux. With unity Lewis number assumption, the mass diffusivity $D_m$ is calculated through $D_m = k/\rho C_p$. Moreover, $\dot{\omega}_m$ is the production or consumption rate of $m$-th species by all $N$ reactions, and can be calculated from the reaction rate of each reaction $\omega_{m,j}^o$, i.e.

$$\dot{\omega}_m = MW_m \sum_{j=1}^{N} \omega_{m,j}^o. \tag{6}$$

Also, the term $\dot{\omega}_T$ in Eq. (3) accounts for the heat release from chemical reactions and is estimated as $\dot{\omega}_T = -\sum_{m=1}^{M} \dot{\omega}_m \Delta h_{f,m}^o$. Here $\Delta h_{f,m}^o$ is the formation enthalpy of $m$-th species.

The Lagrangian method [19] is used to track the liquid fuel droplets. The equations of mass, momentum, and energy for single droplets are

$$\frac{dm_d}{dt} = -\dot{m}_d, \tag{7}$$

$$\frac{d\mathbf{u}_d}{dt} = \frac{\mathbf{F}_d + \mathbf{F}_p}{m_d}, \tag{8}$$

$$c_{p,d} \frac{dT_d}{dt} = \frac{\dot{Q}_c + \dot{Q}_{lat}}{m_d}, \tag{9}$$

where $m_d = \pi \rho_d d^3/6$ is the mass of a single droplet, where $\rho_d$ and $d$ are the droplet material density and diameter, respectively. $\mathbf{u}_d$ is the droplet velocity vector, $c_{p,d}$ is the droplet heat capacity, and $T_d$ is the droplet temperature. Infinite thermal conductivity of the droplet is assumed since the corresponding Biot number is small in our simulations.

The evaporation rate of the droplet $\dot{m}_d$ is calculated with Abramzon and Sirignano model [20]. Its accuracy in prediction of droplet evaporation in elevated ambient pressures and temperatures has been validated in our recent work [17]. It reads

$$\dot{m}_d = \pi d \rho_f D_f \widetilde{Sh} \ln(1 + B_M), \tag{10}$$

where $\rho_f = p_s MW_m / RT_s$ and $D_f = 3.6059 \times 10^{-3} \cdot (1.8T_s)^{1.75} \cdot \frac{\alpha}{p_s \beta}$ are the density and mass diffusivity at the film over the droplet, respectively [17]. $\alpha$ and $\beta$ are the constants related to specific species [21]. $p_s = p \cdot exp\left(c_1 + \frac{c_2}{T_s} + c_3 \ln T_s + c_4 T_s^{c_5}\right)$ is the surface vapor pressure, with $T_s = (T + 2T_d)/3$ being the droplet surface temperature.

In Eq. (8), $\mathbf{F}_d$ is the Stokes drag, which is modelled as $\mathbf{F}_d = \frac{18\mu}{\rho_d d^2} \frac{C_d Re_d}{24} m_d (\mathbf{u} - \mathbf{u}_d)$ [22]. Here $C_d$ is the drag coefficient and estimated using the Schiller and Naumann model [23], and $Re_d \equiv \frac{\rho d |\mathbf{u}_d - \mathbf{u}|}{\mu}$ is the droplet Reynolds number. Also, $\mathbf{F}_p$ is the pressure gradient force and is calculated from $\mathbf{F}_p = -V_d \nabla p$. Here $V_d$ is the volume of a single fuel droplet.

In Eq. (9), $\dot{Q}_c = h_c A_d (T_g - T_d)$ denotes the convective heat transfer between the gas and liquid phases. Here $A_d$ is surface area of a single droplet. $h_c$ is the convective heat transfer coefficient, and estimated using the correlation of Ranz and Marshall [24] through the modified Nusselt number, i.e. $\widetilde{Nu} = 2 + \left[(1 + Re_d Pr)^{1/3} \max(1, Re_d)^{0.077} - 1\right]/F(B_T)$. $Pr$ is the gas Prandtl number, and $B_T$ is the Spalding heat transfer number [20]. Furthermore, $\dot{Q}_{lat}$ in Eq. (9) accounts for the heat transfer caused by the latent heat of evaporation.

Two-way coupling between the gas and liquid phases are considered based on PSI-CELL method, in terms of mass, momentum, energy and species exchanges. Therefore, the source terms for the gas phase equations read ($V_c$ is cell volume and $N_d$ is the droplet number in a CFD cell)

$$S_m = \frac{1}{V_c} \sum_1^{N_d} \dot{m}_d, \tag{11}$$

$$\mathbf{S}_F = -\frac{1}{V_c} \sum_1^{N_d} (-\dot{m}_d \mathbf{u}_d + \mathbf{F}_d), \tag{12}$$

$$S_e = -\frac{1}{V_c}\sum_1^{N_d}[-\dot{m}_d h(T_d) + \dot{Q}_c], \tag{13}$$

$$S_{Y_m} = \begin{cases} S_m & \text{for the liquid fuel species,} \\ 0 & \text{for other species,} \end{cases} \tag{14}$$

where $-\dot{m}_d \mathbf{u}_d$ in Eq. (12) is the momentum transfer due to droplet evaporation. In Eq. (13), $h(T_d)$ is the fuel vapor enthalpy at the droplet temperature. Note that the energy exchange caused by the hydrodynamic forces is not included since it is of secondary importance for dilute spray detonations [25].

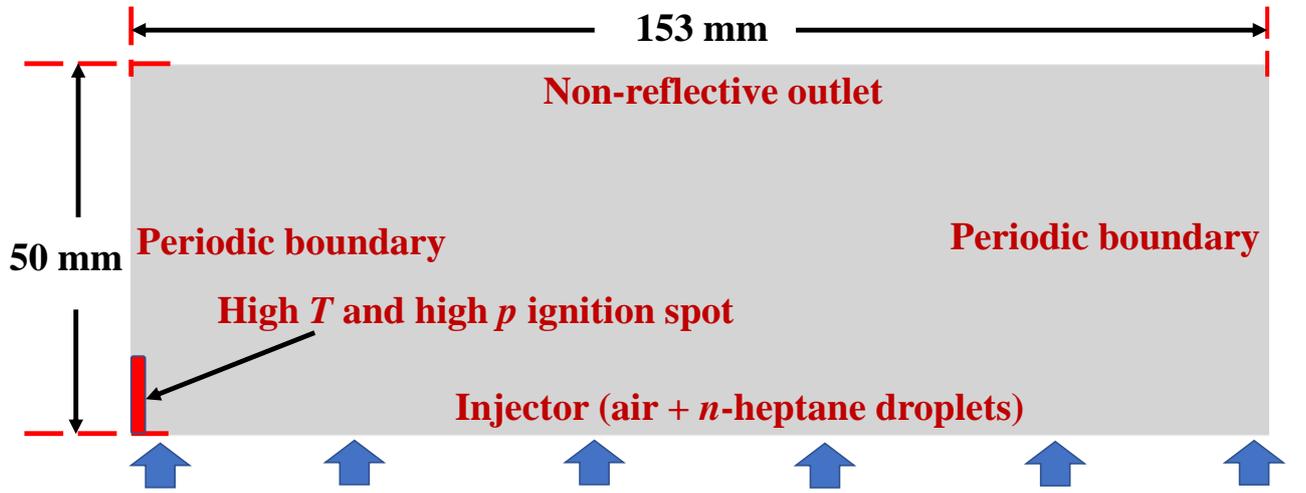

Fig. 1. Computational domain and boundary condition in two-dimensional RDE.

## 2.2 Physical model

Figure 1 shows the schematic of rotating detonation in a two-dimensional (2D) unrolled model RDE chamber. Previous studies about the dimensionality (two or three dimensions) of rotating detonation simulations have confirmed that the results from 2D simulations can well reproduce the key flow and combustion features in RDEs [22,23,26]. In our study, the lengths (*x*-direction) and width (*y*-direction) of the domain are 153 mm and 50 mm, respectively. This extent ensures that the rotating detonation wave and accompanied flow features can be correctly captured.

The boundary conditions of the model RDE chamber are also marked in Fig. 1. Specifically, the outlet is assumed to be non-reflective, which is reasonable since the local flows are supersonic. Periodic boundaries at the left and right sides are enforced, such that the RDW can continuously propagate across the flattened domain.

Through the continuous injectors at the top head in Fig. 1, the spherical droplets of liquid *n*-heptane sprays are injected into the domain with carrier gas, heated air, with the same strategy used by Meng et al. [15,16]. The initial temperature of the *n*-heptane droplets is 323 K to promote rapid evaporation of the droplet. The initial material density of the *n*-heptane droplets is 680 kg/m³. The initial temperature and pressure of the carrier gas air are 700 K and 30 atm, respectively. The liquid equivalence ratio can be varied by changing volume fractions of the liquid fuel droplets in the carrier gas. Moreover, a high-temperature and high-pressure spot (2,000 K and 20 atm) of 1 mm × 12 mm is used in the lower left corner of the combustor, as shown in Fig. 1, to initiate the detonation wave.

It is well known that in practical RDEs [9-11], liquid fuel sprays are always polydispersed, and the size of the fuel droplets are therefore distributed. Different from our previous work [15-17], the effects of the initial droplet diameters of a polydisperse sprays on RDW propagation, in-chamber reactant mixing and propulsion performance are studied. However, to pinpoint the foregoing effects, bi-dispersed droplets with a specified mass ratio are considered in the current study, i.e., one class of fuel droplets with smaller diameter $d_s^0$, whilst the other class with larger sizes $d_l^0$. Their mass ratios are parameterized by liquid equivalence ratios, i.e., $\phi_s$ and $\phi_l$, respectively. They are defined as the mass ratio of the liquid droplets (with $d_s^0$ and $d_l^0$) to the carrier gas air from the injector. In all the simulations, the total liquid fuel equivalence ratios, i.e., $\phi_t = \phi_s + \phi_l$, are fixed to be unity. As such, varying either of the equivalence ratio, $\phi_s$ or $\phi_l$, would lead to change of the other. The initial diameter of the smaller droplet class $d_s^0$ is fixed to be 2 μm in all our cases, whereas $d_l^0$ varies from 5 to 20 μm.

Moreover, in this study, pure *n*-heptane sprays with *in-situ* evaporation in the RDE model combustor will be considered, i.e., no pre-vaporization effects. Therefore, this is closer to the practical RDE implementations. In published literature, very limited work has been reported on modelling of pure spray RDE, except the recent one by Ren and Zheng [14], where pure kerosene is used as the propellant.

2.3 *Numerical implementation*

The governing equations for both gas and liquid phases are solved by a multiphase reacting flow code *RYrhoCentralFoam* [28,29], which is developed based on a density-based compressible flow solver *rhoCentralFoam* in OpenFOAM 6.0 [29]. Detailed validations and verifications show that the solver can accurately predict shock waves, species diffusion, shock-chemistry interaction, detonation cell size and frontal structure [28,29,31]. It has been successfully used for modelling detonative combustion with gaseous and liquid fuels [15,18], as well as supersonic combustion [29,32,33].

The cell-centered finite volume method is used to discretize the gas phase equations, i.e., Eqs. (1)−(4). The second-order implicit backward scheme is used for time marching of the gas phase variables. The time step is about $10^{-9}$ s, which leads to a maximum Courant number of 0.1. Moreover, second-order Godunov-type upwind-central scheme is used to calculate the convection terms in the momentum equations. The total variation diminishing scheme is applied for the convection terms in the energy and species mass fraction equations.

Two-step chemical mechanism for *n*-heptane is used in this work, which includes six species (i.e., *n*-$C_7H_{16}$, $CO$, $CO_2$, $H_2O$, $O_2$, $N_2$) and two reactions. The chemical mechanisms are listed in Table 1 with their respective parameters for Arrhenius kinetics. This mechanism has been validated against a detailed mechanism [33] and the results show that it can correctly reproduce the detonation propagation speed, pressure, and temperature at both von Neumann and Chapman–Jouguet (C-J) points in the ZND (Zeldovich−von Neumann−Döring) structures corresponding to a wide range of operating conditions [16]. The two-step chemistry is deemed sufficient in this work since detailed gaseous chemistry is not focused on here; instead, we are more interested in detonation propagation speed, overall propulsion performance and droplet dynamics in liquid fueled RDE.

Table 1. Chemical mechanism for $n$-$C_7H_{16}$ combustion (units in cm-sec-mole-cal-Kelvin). $A$ is the pre-exponential factor, $n$ is the temperature exponent, $E_a$ is the activation energy, $a$ and $b$ are the fuel and oxidizer reaction orders, respectively.

|   | Reaction | $A$ | $n$ | $E_a$ | $a$ | $b$ |
|---|---|---|---|---|---|---|
| I | $2n$-$C_7H_{16}+15O_2 \Rightarrow 14CO+16H_2O$ | $6.3 \times 10^{11}$ | 0.0 | 30,000.0 | 0.25 | 1.5 |
| II | $2CO+O_2 \Leftrightarrow 2CO_2$ | $4.5 \times 10^{10}$ | 0.0 | 20,000.0 | 1.0 | 0.5 |

For the liquid phase, the Lagrangian equations, i.e., Eqs. (7)−(9), are solved with the first order Euler method. With the PSI-CELL implementations, two-way coupling between the gas and liquid phases about species, mass, momentum, and energy exchanges is performed for each time step, through Eqs. (11)−(14). The droplet breakup model by Reitz [34] is used, which can accurately simulate the break-up of droplets under engine relevant conditions and also successfully used for spray detonation modelling [17].

The computational domain in Fig. 1 is discretized with uniform 496,000 Cartesian cells for the Eulerian flow field calculations and the cell spacing size is 125 μm. Mesh sensitivity analysis is also performed, which demonstrates that further refinement of the mesh would not change the predicted detonation speed and key features of the rotating detonative flow fields. Additionally, in the hybrid Eulerian−Lagrangian method with point-force approximation, the Lagrangian droplet diameter should be smaller than the Eulerian cell size [35]. This is because the gas phase quantities near the droplet surfaces (critical for estimating the two-phase coupling, e.g. evaporation) can be well approximated using the interpolated ones at the location of the sub-grid droplet [36]. In our simulations, the ratio of the Eulerian cell size and Lagrangian droplets, $\theta$, range from 6.25 to 62.5, which is well above or close to the criterion, $\theta > 10$, as suggested by Sontheimer et al. [37] and Luo et al. [38]. As such, the current Eulerian mesh resolution is expected to be sufficient for capturing the flow field, droplet dynamic behaviors and gas−liquid bi-directional coupling in liquid fuel rotating detonations.

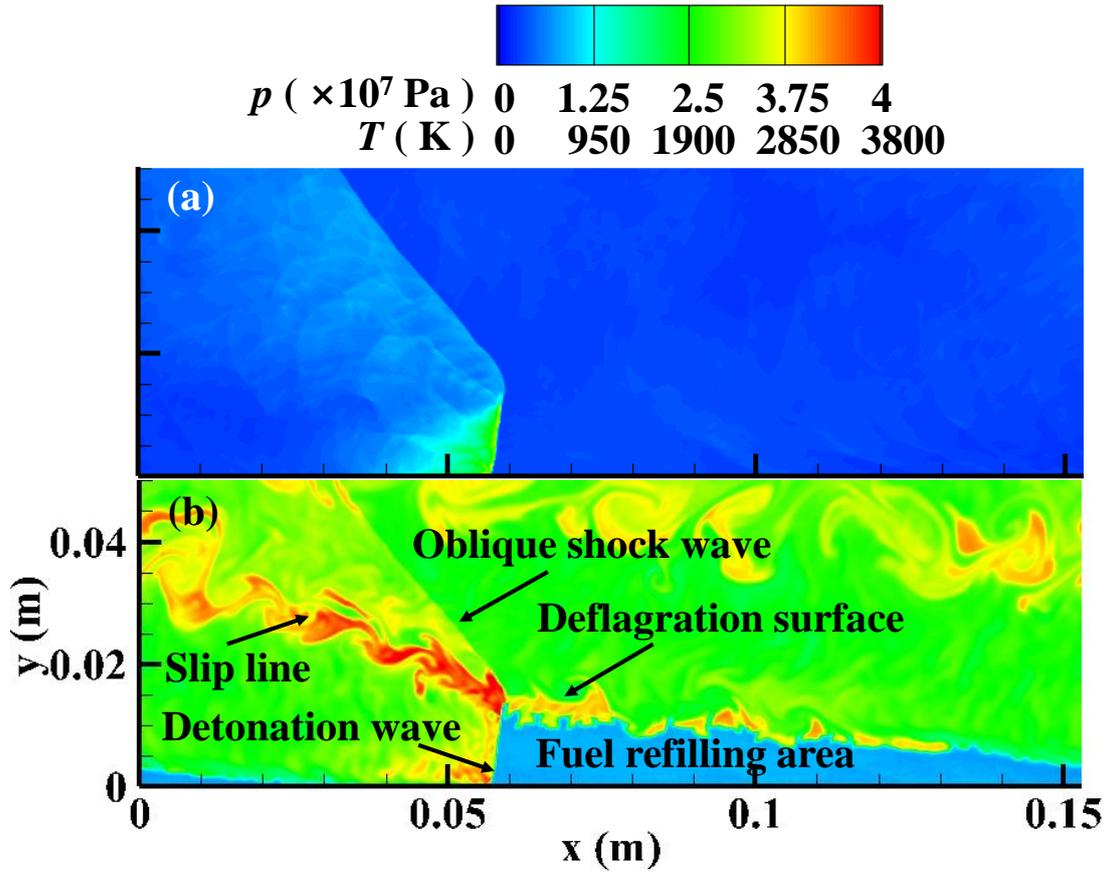

Fig. 2. Contours of (a) pressure and (b) gas temperature. $d^0 = 2$ μm and $\phi_t = 1.0$.

## 3. Results and discussion

### 3.1 *RDW propagation in fuel sprays*

The features of rotating detonations in sprayed *n*-heptane fuels will be demonstrated in this section. Three cases are considered: (1) mono-sized sprays with initial droplet diameter $d^0 = 2$ μm; (2) mono-sized sprays with $d^0 = 10$ μm; and (3) bi-disperse sprays with 50% droplets of $d_s^0 = 2$ μm and 50% large droplets of $d_l^0$ = 10 μm. Be reminded that the (total) liquid fuel equivalence ratios in these three cases are identical, i.e. $\phi_t$ = 1.0. The key information about the gas phase and liquid phase is listed in Table 2.

Table 2. Information about the gas phase and liquid phase in cases 1−3. $T_0$ and $p_0$ are total temperature and total pressure of carrier air, $\phi_t$ is total liquid fuel equivalence ratio, $T_d^0$ is temperature of droplets, $d^0$ is mono-sized sprays with initial droplet diameter, $d_s^0$ and $d_l^0$ are initial droplet diameter of small droplets and large droplets in bi-disperse sprays.

| Case | Gas phase | | | Liquid phase | | | |
|---|---|---|---|---|---|---|---|
| | $T_0$ (K) | $p_0$ (atm) | $\phi_t$ | $T_d^0$ (K) | $d^0$ (μm) | $d_s^0$ (μm) | $d_l^0$ (μm) |
| 1 | 700 | 30 | 1 | 323 | 2 | - | - |
| 2 | | | | | 10 | - | - |
| 3 | | | | | - | 2 | 10 |

Figure 2 shows the contours of pressure and gas temperature corresponding to case 1. The results are extracted after the RDW runs over ten cycles. In this work, one cycle means that the RDW propagates from the left periodic boundary to the right one. The key features of rotating detonation flow field, including detonation wave, oblique shock wave, slip line and deflagration surface, are well predicted, as marked in Fig. 2(b). The triangular fuel refilling area is generally regular, and thereby liquid fuel evaporation and fuel vapor / oxidizer mixing can proceed therein. The average detonation propagation velocity of the detonation wave under the current condition is about 1,760 m/s, which is lower than the purely gaseous RDW speed (1,830 m/s) under the same pressure and total temperature conditions. The C-J speed in the corresponding gaseous conditions is 1,835.7 m/s. As such, the velocity deficits are 3.8% and 4.1%, respectively.

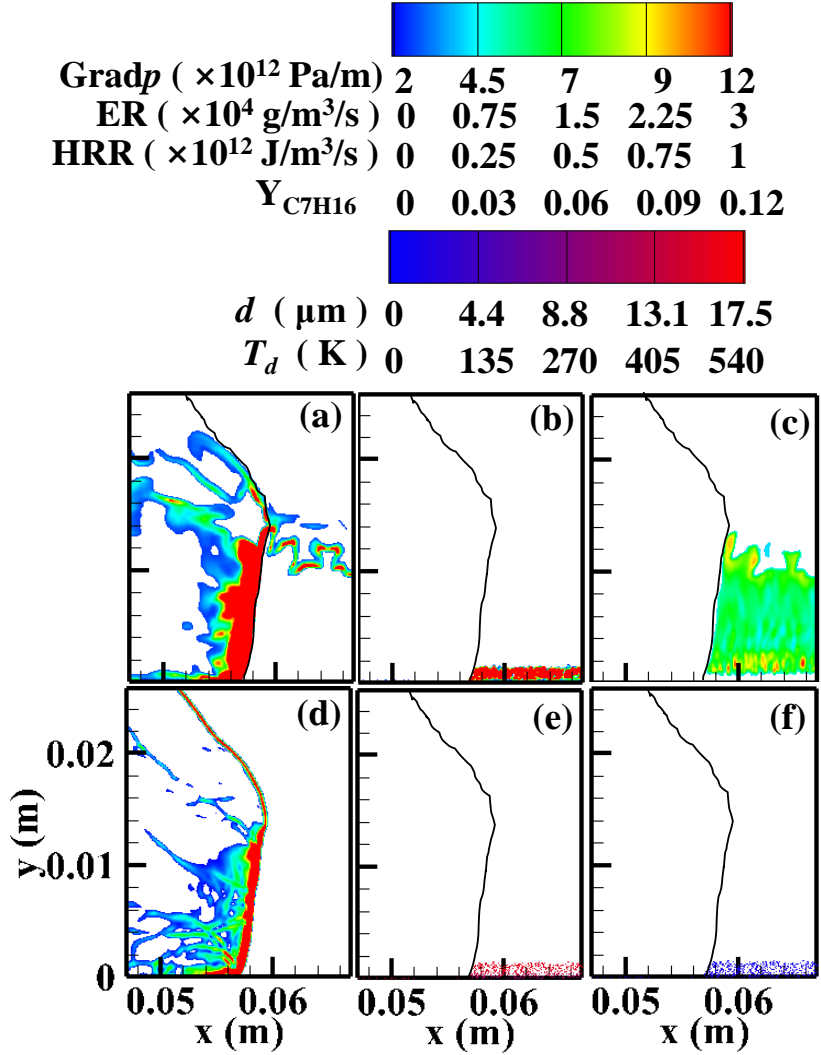

Fig. 3. Contours of (a) heat release rate, (b) evaporation rate, (c) *n*-heptane vapor mass fraction, (d) pressure gradient magnitude, (e) droplet temperature and (f) diameter. $d^0 = 2$ μm and $\phi_t = 1.0$. Solid line: detonation and oblique shock waves.

Figure 3 further shows the enlarged views about the distributions of Heat Release Rate (HRR), Evaporation Rate (ER), *n*-$C_7H_{16}$ vapor mass fraction, pressure gradient magnitude, Lagrangian droplet temperature and diameter near the detonation wave in Fig. 2. Note that in Fig. 3(b) the evaporation rate is the volumetric source term $S_m$ in Eq. (11) and therefore it is a Eulerian quantity. One can see from Fig. 3(a) that high heat release rate can be found along the detonation wave, except near the triple point. There, the leading shock (solid line) and reaction front (with high HRR) are decoupled. This is because the fuel vapor ahead of it has been consumed by the deflagration surface.

After being injected into the combustor, the *n*-heptane droplets are quickly heated close to the saturation temperature (see Fig. 3e, about 540 K) and then start to vaporize quickly and therefore considerable evaporation can be observed near the injector with fast reduction of the droplet size, demonstrated in Figs. 3(b) and 3(f). The height of the evaporating droplet layer is small, about 1.5 mm, beyond which no droplets exist. In the fuel filling area, the resultant *n*-heptane vapor mass fraction is close to stoichiometry (about 6.02%, see Fig. 3c), indicating the complete evaporation of the liquid fuels. One can also see from Fig. 3(c) that the fuel vapor mass fraction is relatively uniform ahead of the RDW, which implies the efficient mixing of the fuel vapor and oxidizer inside the refilling area. Moreover, it is shown from Figs. 3(c), 3(e) and 3(f) that there are no *n*-heptane droplets behind the detonation wave, and therefore all the fuels have been consumed by the rotating detonation wave or deflagration surface.

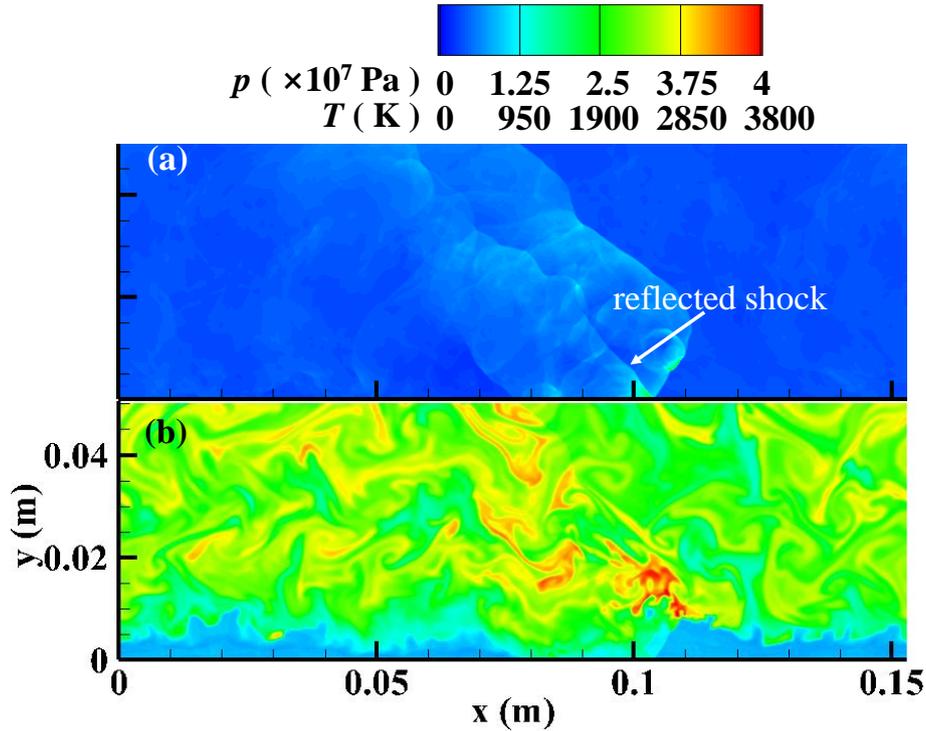

Fig. 4. Contours of (a) pressure and (b) gas temperature. $d^0 = 10$ μm and $\phi_t = 1.0$.

Figure 4 shows the contours of pressure and gas temperature in case 2, in which the initial droplet diameter $d^0$ is increased to 10 μm. In this case, the detonation wave is quenched after propagating after about 5 cycles. The transient extinction process will be discussed in detail in Section 3.2. Briefly, the fuel refilling area becomes less organized, compared to that in case 1. No pronounced temperature rise is observed along the interface between the fuel refilling area and burned product gas. This indicates that less deflagrative combustion occurs due to insufficient fuel vapor. Moreover, the leading shock wave becomes oblique and is reflected at the inlet. Since the pressure immediately behind the RDW is higher than the total pressure, based on our gas injection method [15,16], it is assumed to be a solid wall. This reflected shock is almost parallel to the oblique shock connected with the leading shock.

Figure 5 shows the distributions of HRR, ER, $n$-$C_7H_{16}$ vapor mass fraction, pressure gradient magnitude, Lagrangian droplet temperature and diameter around the detonation wave corresponding to the same instant in Fig. 4. The HRR contour in Fig. 5(a) shows that the detonative combustion only proceeds behind a small fraction of the leading shock, roughly corresponding to the downstream ($y >$ 0.006 m) of the fuel refilling area. When $y <$ 0.006 m, finite distance between the leading shock wave and reaction front can be seen, and therefore no detonations occur there. However, one can find that a Secondary Rotating Detonation Wave (SRDW) exists near the injector. This phenomenon is also reported by Ren and Zheng [14] in liquid kerosene RDE. The formation of SRDW can be attributed to: (1) existence of the reflected shock wave; (2) sufficient $n$-heptane vapor ahead of the reflected shock wave (behind the leading shock). The second reason can be more clearly shown in Figs. 5(b) and 5(c), through which high evaporation rate and fuel vapor concentration can be found between the reflected and leading shocks. How the secondary rotating detonation wave evolves during a detonation extinction process will be further interpreted in Section 3.2. In this case, the height of the evaporating droplet distribution zone is much higher than that in case 1, because larger droplets may have longer heating and evaporation timescales.

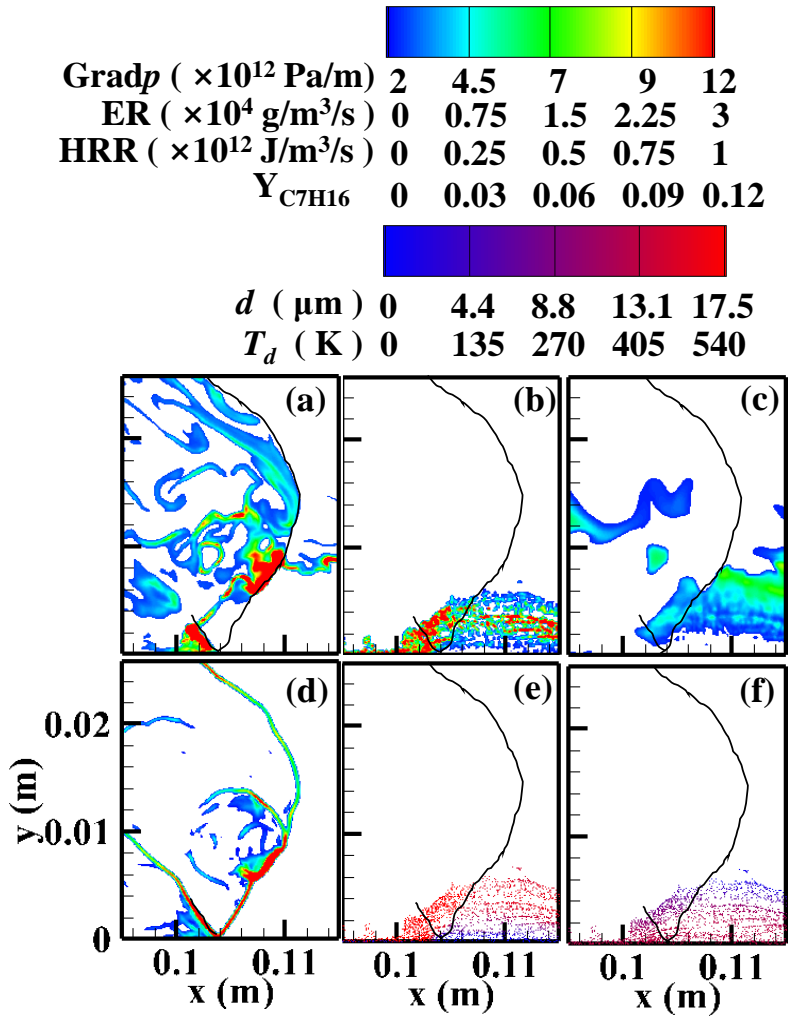

Fig. 5. Contours of (a) heat release rate, (b) evaporation rate, (c) *n*-heptane vapor mass fraction, (d) pressure gradient magnitude, (e) droplet temperature and (f) diameter. $d^0 = 10$ μm and $\phi_t = 1.0$. Solid line: detonation and oblique shock waves.

Plotted in Fig. 6 are the contours of pressure and gas temperature in the bidisperse sprays, i.e., case 3. Similar to the results of case 2 in Fig. 5, the leading shock wave is inclined, and a reflected shock wave is present. However, different from case 2, case 3 is characterized by continuously rotating detonation propagation across the model RDE chamber. Although the RDW are stable both in cases 1 and 3, nevertheless, the morphology of the RDW is different, which can be more clearly seen in Fig. 7.

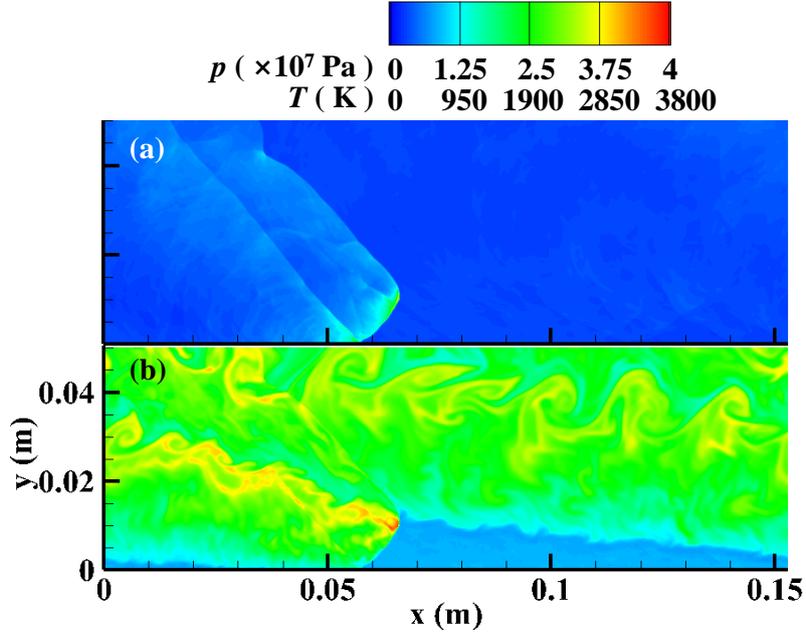

Fig. 6. Contours of (a) pressure and (b) gas temperature. $d_s^0 = 2$ μm (50%), $d_l^0 = 10$ μm (50%), $d_s^0 = 2$ μm (50%) and $\phi_t = 1.0$.

Figure 7 shows zoomed contours of the HRR, ER, $n$-$C_7H_{16}$ vapor mass fraction, pressure gradient magnitude, Lagrangian droplet temperature and diameter near the detonation wave in Fig. 6. More heat release behind the leading shock wave can be found in Fig. 7(a), compared to the counterpart results in Fig. 5(a). This can confirm the effects of the small droplets in fuel vapor supply and hence sustain the detonative combustion. Likewise, local extinctions of the detonation wave can be also observed near the injector in Fig. 7(a). A SRDW along the reflected shock wave is also present, which is the same as that in Fig. 5(a). The average detonation propagation speed is about 1,750 m/s, slightly lower than that in Fig.2. In Fig. 7, one can also see that the fuel droplets are dispersed almost in the entire fuel refilling area, and this is because 50% of the fuel sprays have larger diameter (10 μm), which have longer heating and evaporation time in the fuel refilling area.

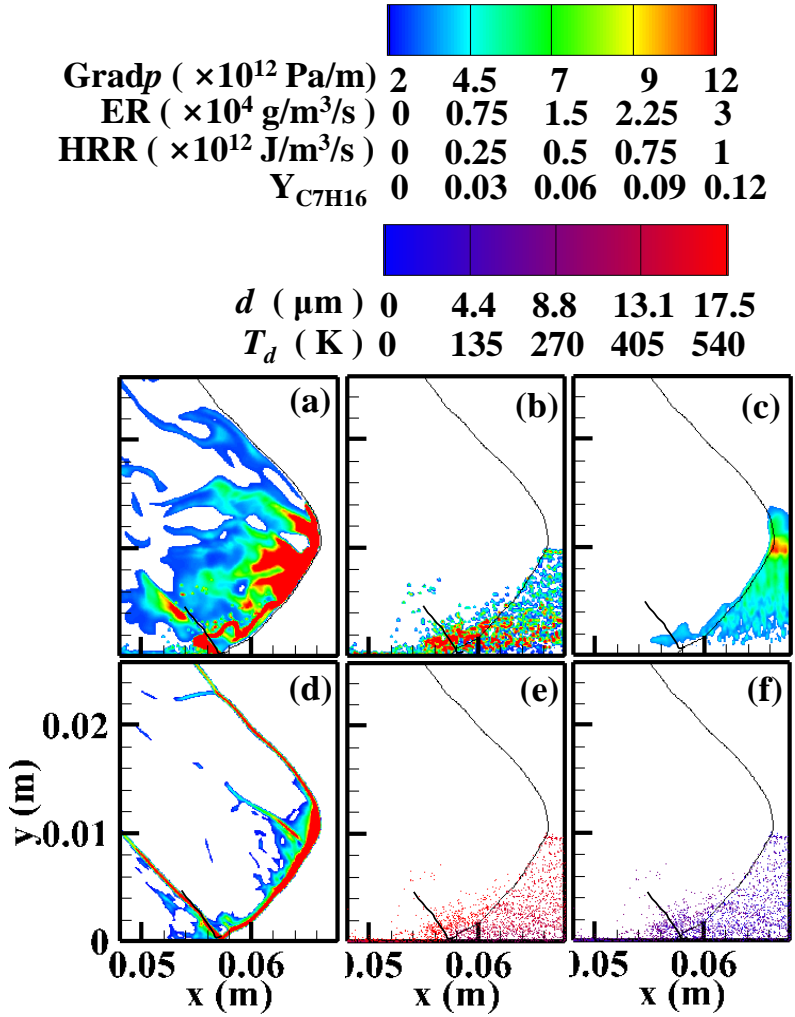

Fig. 7. Contours of (a) heat release rate, (b) evaporation rate, (c) *n*-heptane vapor mass fraction, (d) pressure gradient magnitude, (e) droplet temperature and (f) diameter. $d_s^0 = 2$ μm (50%), $d_l^0 = 10$ μm (50%) and $\phi_t = 1.0$. Solid line: detonation and oblique shock waves.

## 3.2 *RDW extinction in fuel sprays*

It has been shown from case 2 that the rotating detonative combustion fueled with *n*-heptane sprays are quenched after propagating about five cycles. Their transient will be further discussed in this section, about the how the main and secondary rotating detonation wave evolve. Figure 8 demonstrates the time sequences of pressure and gas temperature during the detonation extinction process. At 1,340 μs (same as that in Fig. 4), the RDW still exists. From 1,360 μs to 1,380 μs, the height of the detonation wave gradually decreases. Moreover, since the pressure behind the detonation wave is higher than the total pressure of the inlet air, the fuel sprays cannot be injected into the combustor, which leads to a gradually reduced fuel filling area. From 1,400 μs to 1,440 μs, the RDW gradually becomes weak, which can be confirmed by the decreased temperature and pressure near the detonation wave. Eventually, the detonation wave is extinguished.

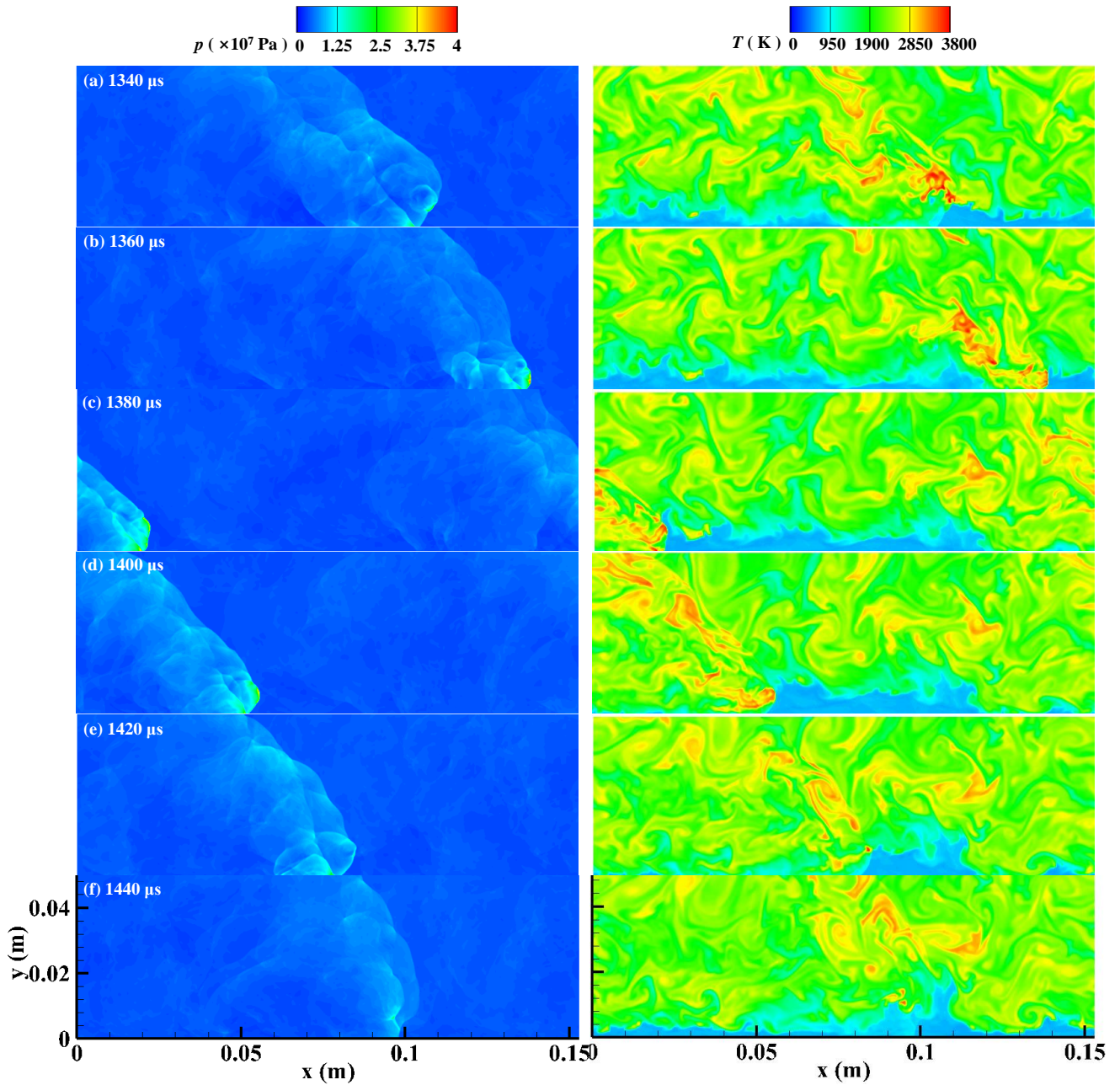

Fig. 8. Extinction process of a detonation wave in *n*-heptane sprays. $d^0$ = 10 μm and $\phi_t$ = 1.0.

Figure 9 shows the evolutions of the HRR and ER corresponding to the above detonation extinction process. Figures 9(a)−9(f) correspond to the six instants in Fig. 8 (1,340 μs – 1,440 μs). It can be found that the detonation wave undergoes an extinction and re-ignition process. Specifically, at 1,340 μs, due to the large droplet diameter, the droplets are unable to evaporate completely in the fuel refilling area and a large amount of evaporating droplets exist after the detonation wave. These droplets result in a high evaporation rate in this region. There is a significant discontinuity in the heat release rate on the detonation wave. The detonation wave experiences the first instantaneous extinction.

From Fig. 9(a) to 9(b), although the detonation wave is extinguished, the higher temperature after the wave allows the droplets to continue to evaporate and eventually cause the detonation wave to re-ignite. After that, the heat release behind the leading shock is more distributed, indicating the enhanced detonative combustion, as shown in Fig. 9(b)-(d). Meanwhile, the number of evaporating droplets after the detonation wave gradually increases during this process and eventually leads another severe localized extinction of detonation combustion behind the leading shock, as shown in Fig. 9(e). Finally, another extinction at 1440 μs can be found in Fig. 9(f). Eventually, the pressure wave is fully decoupled from the combustion wave and the detonation wave is extinguished. Moreover, the height of the secondary detonation wave from the reflected shock wave is low at 1,340 μs, about 2 mm (see Fig. 9a). From 1,360 to 1380 μs, as the detonation wave is reignited and gradually develops, the height of the secondary detonation wave increases to 5 mm. The secondary detonation wave from the reflected shock also becomes quenched and at 1420 μs, it is no longer observable in Fig. 9(f).

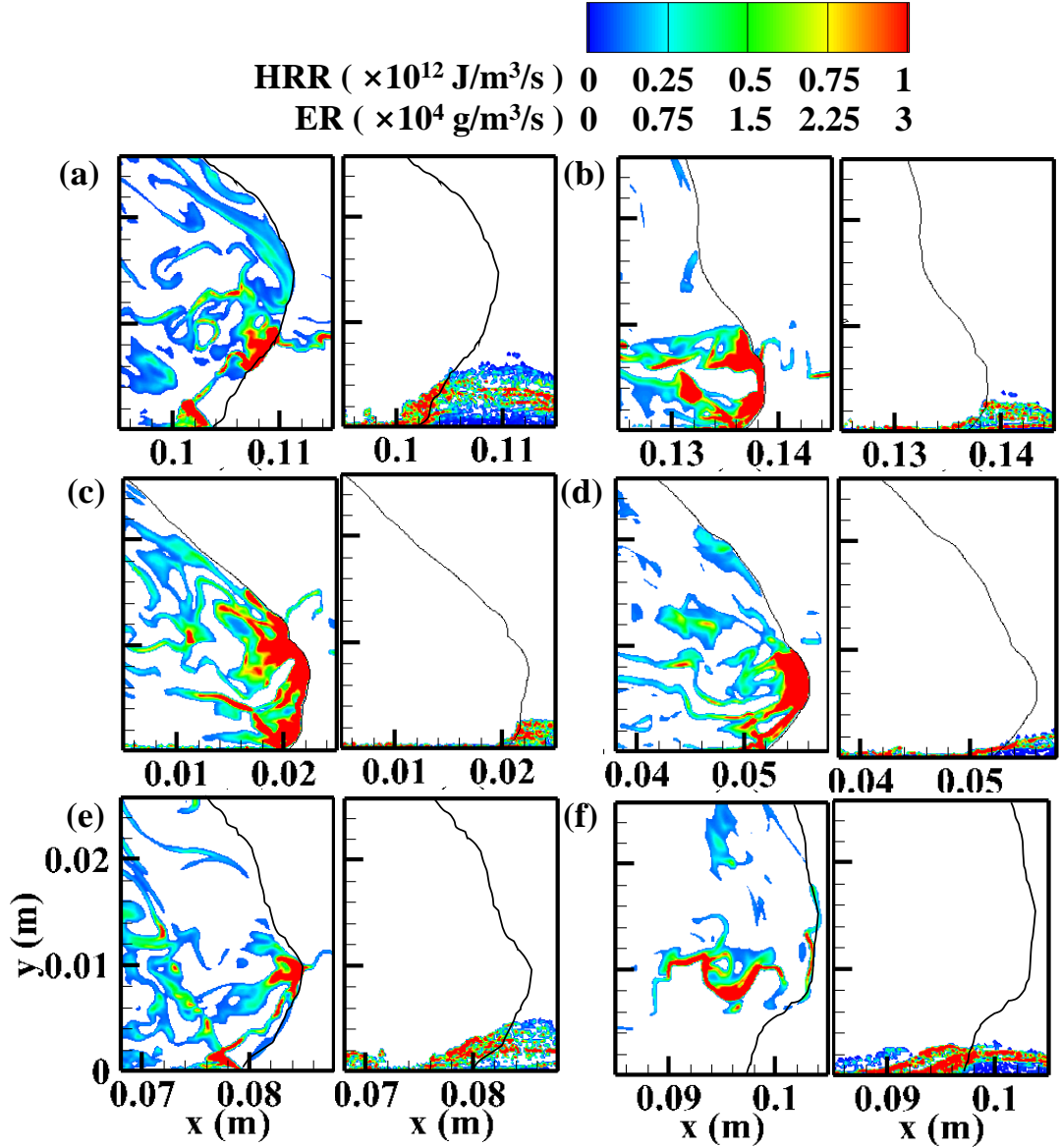

Fig. 9. Time sequence of heat release rate (left column) and evaporation rate (right column) in a detonation extinction process in case 2. Solid line: detonation and oblique shock waves.

### 3.3 *Reactant mixing and detonated fuel fraction*

The structure and extinction process of RDW with mono-sized sprays have been discussed in Sections 3.1 and 3.2. In Section 3.3, the effects of larger droplet diameter ($d_l^0$) in the bi-disperse sprays on reactant mixing, effective equivalence ratio and detonated fuel fraction will be investigated. Figure 10 shows the contours of *n*-heptane vapor mass fraction and equivalence ratio from the cases of $d_l^0 =$ 5, 7.5 and 10 μm, respectively. $\phi_l = 0.5$ and $d_s^0 = 2$ μm. In this analysis, the effective equivalence ratio $\phi_{eff}$ is defined as the ratio of required stoichiometric oxygen atoms to the available oxygen atoms [39]. The former is defined as the minimum number of oxygen atoms demanded to convert all carbon and hydrogen atoms to $CO_2$ and $H_2O$, respectively [39], i.e.,

$$\phi_{eff} = \frac{n_C + n_H/4}{n_O/2}, \quad (15)$$

where $n_C$, $n_H$, and $n_O$ denote the number of available carbon, hydrogen and oxygen atoms, respectively. The reader should be reminded that since it is based on element conservation, $\phi_{eff}$ is also well defined in the detonation product area. However, the ones in the un-detonated mixtures (such as triangular fuel refilling area) are most relevant for our analysis. Also, only the atoms in the gas phase are considered and no contribution (such as hydrogen or carbon atoms) from the liquid fuels is included.

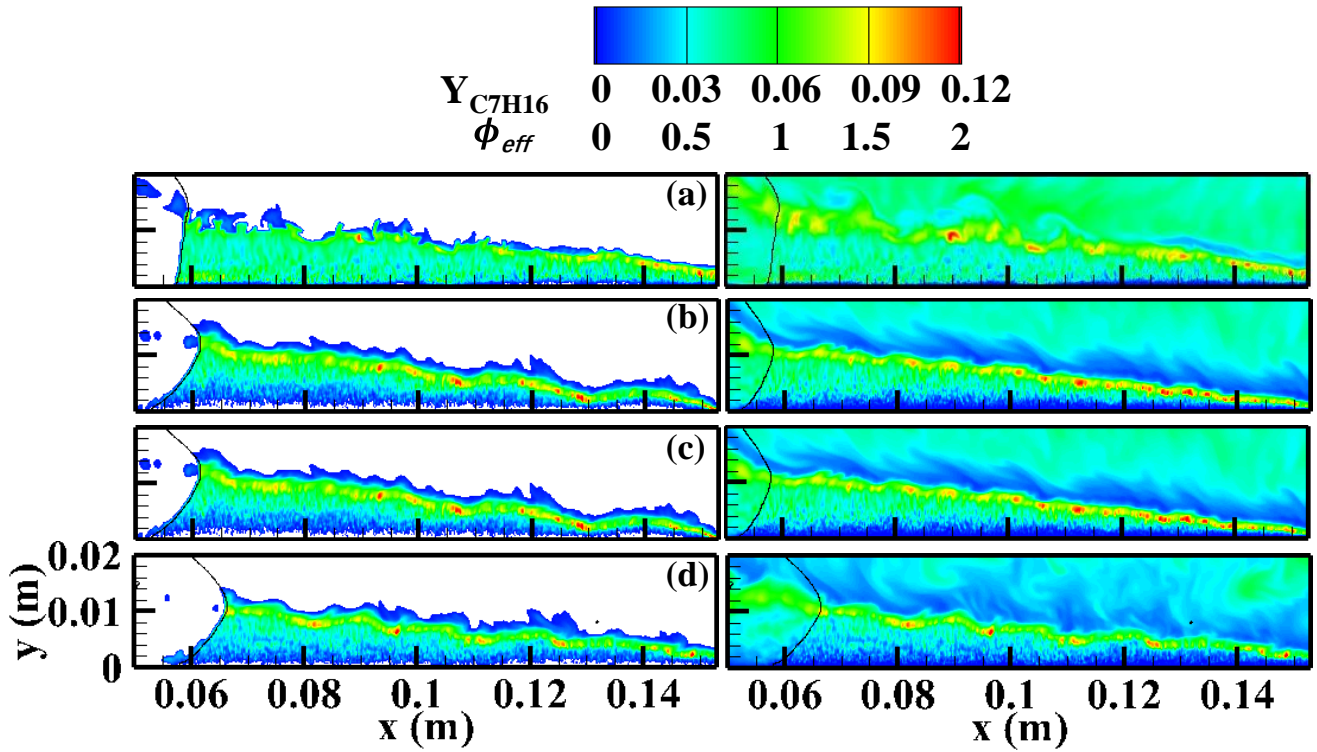

Fig. 10. Contours of (left column) *n*-heptane vapor mass fraction and (right column) effective equivalence ratio: (a) $d_l^0$ = 2 μm, (b) $d_l^0$ = 5 μm, (c) $d_l^0$ = 7.5 μm and (d) $d_l^0$ = 10 μm. $\phi_l$ = 0.5 and $d_s^0$ = 2 μm. Solid line: detonation and oblique shock waves.

One can see from Fig. 10 that, as $d_l^0$ increases from 2 to 10 μm, less vapor is released from the sprayed droplets intermediately after they are injected into the RDE chamber, resulting in less distributions of *n*-$C_7H_{16}$ vapor near the top heat injectors. As the droplets gradually evaporate downstream of the fuel refilling area, gradual increase of the mass fraction of $n$-$C_7H_{16}$ vapor can be observed. This is particularly true for $d_l^0 \geq 5$ μm in Figs. 10(b)-10(d). The difference in droplet diameter can lead to a significant difference in the time required for complete evaporation of the droplet. The difference in the distribution of $n$-$C_7H_{16}$ vapor in the fuel refilling area further results in a change of the angle of the detonation wave. As shown in Fig. 10, the detonation wave propagates in a stratified reactant mixtures along the detonation wave height direction, from fuel-lean, to stoichiometric, to spotty fuel-rich compositions along the deflagrative contact surface. Furthermore, in the fuel refilling area, the distance in the *y*-direction where the equivalence ratio reaches unity increases with the $d_l^0$, which ultimately leads to an increase in the angle of the detonation wave with the size of the larger droplet class $d_l^0$. *n*-Heptane vapor is accumulated near the contact surface between the refilled fuel and detonated product, particularly obvious in Figs. 10(b)−10(d). The reason for this peculiar phenomenon and its effects on rotating detonations have been explained in Ref. [16].

As shown in Fig. 10, the similar tendencies of the *n*-heptane vapor inside the fuel refilling area can also been observed from the distributions of the effective equivalence ratio. For $d_l^0 = 2$ μm, overall unity equivalence ratio in the fuel refilling area is found, indicating the fast evaporation and efficient vapor/oxidizer mixing before the detonation wave arrives. However, for the rest cases, the equivalence ratio is almost zero near the top head, and gradually increases towards unity along the *y*-direction. This is consistent with the results of fuel vapor distributions discussed above. At the fuel-product contact surface, locally rich pockets can be found, with $\phi > 2.0$.

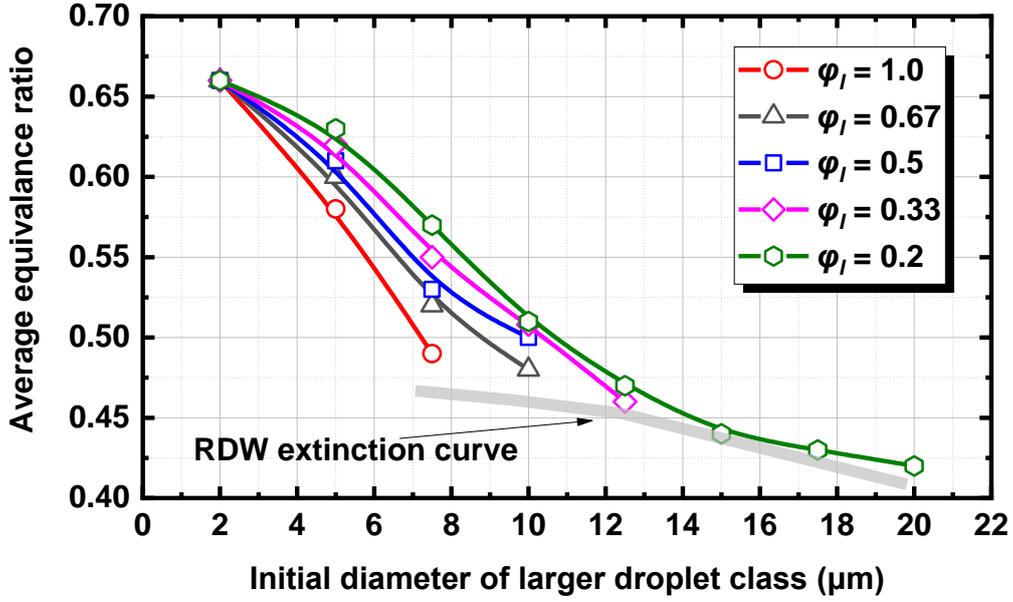

Fig. 11. Average equivalence ratio in the fuel refilling area as a function of the diameter of larger droplet class. $d_s^0 = 2$ μm and $\phi_t = 1.0$.

Figure 11 further quantifies the average equivalence ratio as a function of the diameter $d_l^0$ of the larger fuel droplet class. Different liquid fuel equivalence ratios $\langle\phi_{eff}\rangle$ for larger droplet class are considered, i.e., $\phi_l = 0.2-1.0$. Here the averaging is performed based on the fuel refilling area based on ten uncorrelated time instants. Since the droplet evaporation is very limited when they are first injected into the combustor, equivalence ratio around the inlet is almost zero. This makes the average equivalence ratio $\langle\phi_{eff}\rangle$ in the fuel filling area well below 1, with a maximum value of 0.66. As $d_l^0$ increases, the evaporation rate of the droplets decreases and the area near the inlet with an equivalence ratio close to zero gradually increases. This makes the average equivalence ratio in the fuel filling area gradually decrease.

To further interpret the premixedness of the reactants in liquid fueled rotating detonative combustion, the Flame Index (*FI*) is used here to identify the local combustion regimes, i.e., premixed ( $FI = +1$) or non-premixed ( $FI = -1$) condition [17,40]. It is defined as

$$FI = \frac{\nabla Y_F \cdot \nabla Y_O}{|\nabla Y_F||\nabla Y_O|}, \tag{16}$$

where $Y_F$ and $Y_O$ represent the mass fractions of gaseous *n*-heptane and oxygen, respectively. Figure

12 shows the contours of flame index in the RDE combustor corresponding to the foregoing four cases. As shown in Figs. 12(a) to 12(d), a value of -1 for *FI* is found in the fuel refilling area, which implies that the fuel and oxidizer mixing proceeds there. This is in line with the findings from Ref. [17], although the inter-injector spacing is considered therein. However, the deflagration surface and detonation wave are dominated by premixed combustion (*FI* = +1).

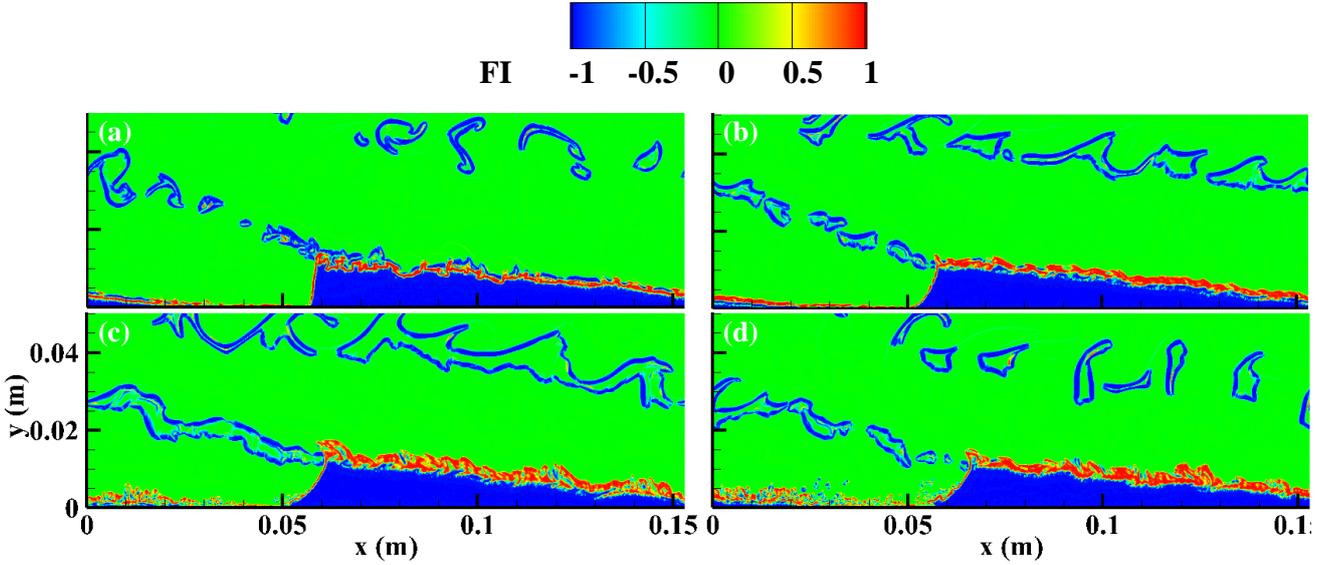

Fig. 12. Contours of flame index in the combustor: (a) $d_l^0$ = 2 μm, (b) 5 μm, (c) 7.5 μm and (d) 10 μm. $\phi_l$ = 0.5 and $d_s^0$ = 2 μm.

The detonated fuel fraction $\psi$ [16,41] is further adopted to measure the percentage of the *n*-heptane fuel burned by the rotating detonation. It can be estimated from

$$\psi = \frac{\int_V \omega t_{C_7H_{16}} dv}{\int_V \omega t_{C_7H_{16}} dv + \int_V \omega f_{C_7H_{16}} dv}, \qquad (17)$$

where $\omega t_{C_7H_{16}}$ and $\omega f_{C_7H_{16}}$ are the volumetric consumption rates of detonated and deflagrated *n*-$C_7H_{16}$, respectively. *V* represents the computational domain. Note that the *n*-heptane fuel is deemed denoted (deflagrated) when the corresponding heat release rate is greater than or approximately equal to (less than) $10^{13}$ J/m³/s [29]. This value is determined from a stand-alone C-J *n*-heptane detonation calculation with The Shock & Detonation Toolbox [41]. When we slightly adjust the foregoing criterion around this numerical value, the obtained detonated fuel fraction is almost not affected. This shows the limited sensitivity of $\psi$ to the HRR criterion.

Figure 13 shows the change of detonated fuel fraction $\psi$ as a function of the diameter $d_l^0$ of the larger fuel droplet class. For comparison, the result of gaseous RDC with $\phi = 1.0$ (i.e., full vaporization before injection) is also shown. The detonated fuel fraction $\psi$ in the simulated liquid fuel RDE are 6%−18% lower than that (0.9) of the corresponding gaseous RDE. Overall, regardless of the larger droplet equivalence ratio $\phi_l$, the detonated fuel fraction $\psi$ monotonically decreases with increased $d_l^0$. This is because as the diameter $d_l^0$ increases, the droplets do not evaporate quickly after being injected into the combustor. There are large amount of fuel droplets crossing the detonation wave and continuing evaporating there. The released vapor mixes with the local oxidizer and is deflagrated (HRR is below $10^{13}$ J/m$^3$/s). As $d_l^0$ increases, the droplets behind the detonation wave increases, which would lead to reduced detonated fuel fraction $\psi$. This trend is observed all the larger droplet equivalence ratio $\phi_l$.

Moreover, as $\phi_l$ increases, from instance, from 0.2 to 1.0, the diameter of the largest droplets with which a stable detonation wave can be sustained gradually decreases. Besides, under the same $d_l^0$ (< 7.5 μm), $\psi$ increases as $\phi_l$ decreases. Decreased $\phi_l$ indicates increased fraction of small droplets ($d_s^0$ = 2 μm) that can be fully evaporated and which can contribute towards the detonative combustion, which will increase $\psi$.

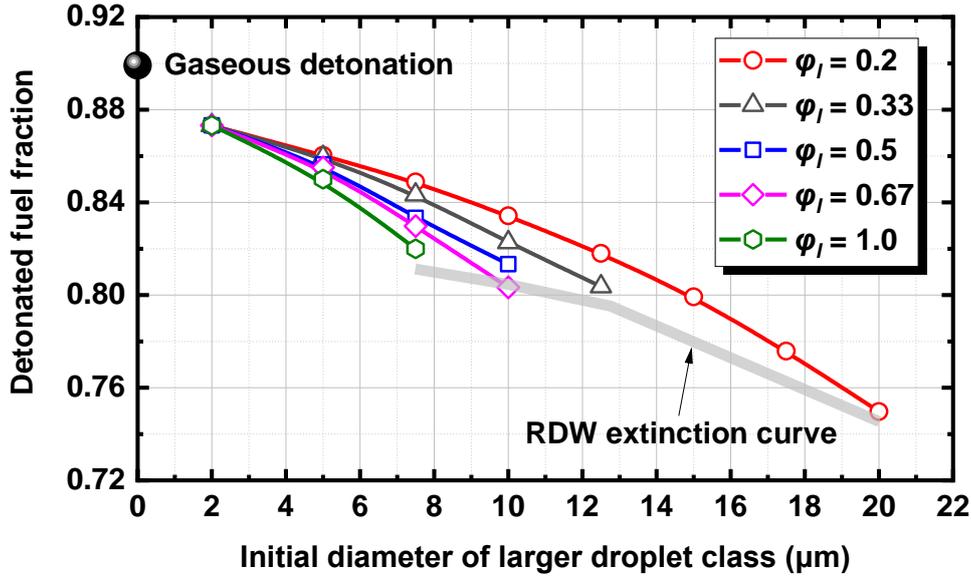

Fig. 13. Detonated fuel fraction as a function of the diameter of larger droplets. $d_s^0 = 2$ μm and $\phi_t = 1.0$.

3.4 *Detonation wave propagation speed*

Figure 14 shows the change of the detonation propagation speed with the diameter of the larger droplets $d_l^0$ in bi-dispersed sprays. The total liquid fuel equivalence ratio $\phi_t$ is 1 for all considered cases. For comparison, the result of gaseous RDC with $\phi_t = 1.0$ (i.e., full vaporization before injection) is also added. It is found that the detonation propagation speeds from liquid fueled RDC are 4%-10% lower than that of the corresponding gaseous RDC. There may be different reasons for the speed deficits, such as nonuniform mixing of oxidizer and fuel in the fuel filling area, and heat or momentum exchange due to the droplets near the detonation front [17]. As $d_l^0$ increases, the detonation propagation speed gradually decreases in all the cases with various $\phi_l$. This is because the larger the size of the fuel droplet, the lower the average equivalence ratio in the fuel filling area, and the propagation velocity of the detonation wave decreases as the equivalence ratio decreases. Under the same equivalence ratio $\phi_l$, the small droplets can release more vapor than the large size droplets, and therefore the detonation propagation speed is higher.

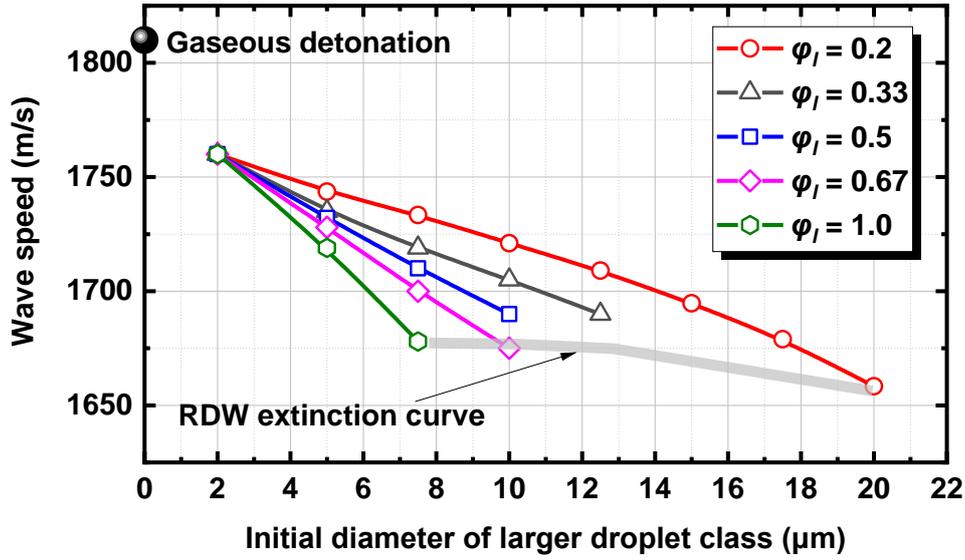

Fig. 14. Detonation wave speed as a function of the droplet diameter. $d_s^0 = 2$ μm and $\phi_t = 1.0$.

The RDW velocity deficit at different $d_l^0$ is given in Figure 15. The velocity deficit is calculated as $\delta v = (v_g - v_d)/v_g$, where $v_g$ is the velocity of gaseous detonation at the same total pressure and temperature, whist $v_d$ is that of two-phase detonations at different $d_l^0$. As shown in Fig. 15, the droplets need to evaporate into gaseous *n*-heptane mixing with air in the fuel refilling area, which makes the distribution of the equivalence ratio in the fuel refilling area non-uniform, as shown in Fig. 10. This non-uniform distribution makes the speed of the detonation wave lower than that of the gaseous detonation under the same conditions. As $d_l^0$ increases, *n*-heptane vapor yield in the fuel refilling area decreases, and accordingly the equivalence ratio in the fuel-refilling area decreases, which makes the detonation wave speed decrease and the velocity deficit increases. Moreover, as the equivalence ratio of lager droplet classes $\phi_l$ increases, the number of droplets increases, the equivalence ratio in the fuel-refilling area decreases, and hence the velocity deficit gradually increases.

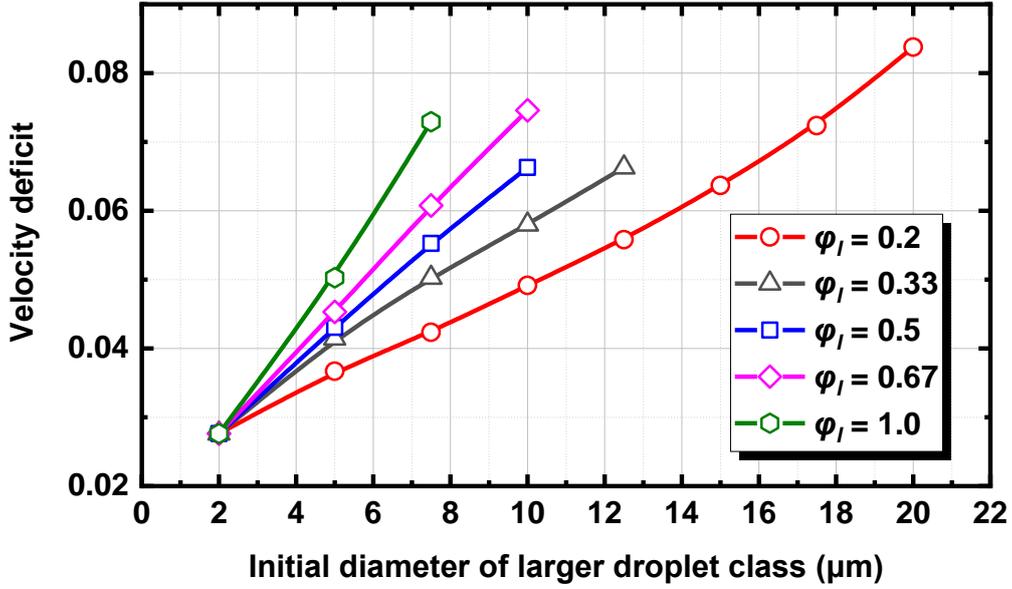

Fig. 15. Velocity deficit as a function of the droplet diameter. $d_s^0 = 2$ μm and $\phi_t = 1.0$.

### 3.5 Propulsion performance

The droplet diameter and spatial distribution in the RDE chamber not only affect the detonation wave propagation, but also the propulsion performance resulting from the detonation combustion. To this end, the specific impulse $I_{sp}$ is calculated

$$I_{sp} = \int_{A_o}[\rho u^2 + (p - p_b)]dA_o/g\dot{m}_F, \tag{18}$$

in which $A_o$ is the area of the outlet, $u$ is the gas velocity at the outlet, $\dot{m}_F$ is the mass flow rate of the fuel, $g$ is gravity acceleration, $p$ is the local pressure at the outlet, and $p_b$ is the backpressure. Figure 16 shows the effect of diameter $d_l^0$ and equivalence ratio $\phi_l$ of the larger droplet class on the specific impulse $I_{sp}$. Specifically, for a given equivalence ratio (such as $\phi_l = 0.5$), as $d_l^0$ increases, the specific impulse gradually decreases. This is because as $d_l^0$ increases, when the droplets are sprayed into the combustor, they cannot quickly evaporate into a gaseous state. This further affects the overall ratio of detonation combustion and hence reduces specific impulse. For a fixed diameter of larger droplet class (such as $d_l^0 = 5$ μm), as $\phi_l$ increases, the number of droplets with a diameter of 2 μm decreases, which means that the number of droplets that can completely evaporate into fuel vapor in the fuel refilling area decreases. The decrease in the number of droplets decreases the average equivalence ratio in the fuel refilling area, which leads to a decrease in specific impulse.

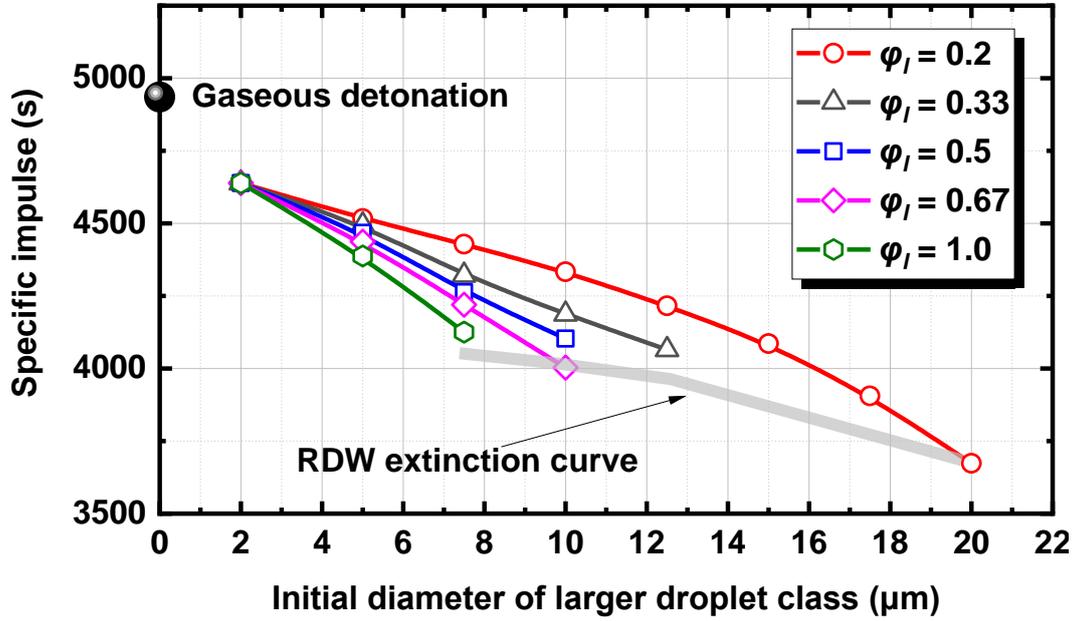

Fig. 16. Specific impulse as a function of droplet diameter. $d_s^0 = 2$ μm and $\phi_t = 1.0$.

The thrust force from the kinetic energy and pressure gain in rotating detonations are also shown in Fig. 17. The thrust from kinetic energy is defined as $F_u = \int_{A_o} \rho u^2 dA_o$, whilst the thrust from pressure gain is $F_p = \int_{A_o} (p - p_b) dA_o$ [17]. The thrust from kinetic energy includes the thrust generated by combustion products as well as by the propellant itself. The thrust generated by combustion products is the thrust from the pressure gain and it is the more important one in the two types of thrust described above, which can be clearly observed in the Fig. 17. The thrust force from pressure gain $F_p$ decreases significantly with the droplet diameter $d_l^0$. As $d_l^0$ increases, the droplet evaporation rate decreases, and the unburned droplets after the detonation wave gradually increases, resulting in deflagration combustion of $n$-heptane vapor near the slip line. This reduces the detonated fuel fraction $\psi$ (see Eq. 17) and eventually leads to a decrease in thrust. For a fixed diameter of larger droplet class (such as $d_l^0 = 5$ μm), as $\phi_l$ increases, the average equivalence ratio $\langle \phi_{eff} \rangle$ in the fuel refilling area decreases, which leads to a decrease in the detonated fuel fraction and ultimately to a decrease in thrust from pressure gain. Since the propellant flow rate is not changed in the calculations of this study, this means that the kinetic energy of the propellant produces almost no change in thrust for different operating conditions, so the trend of $F_u$ is the same as $F_p$.

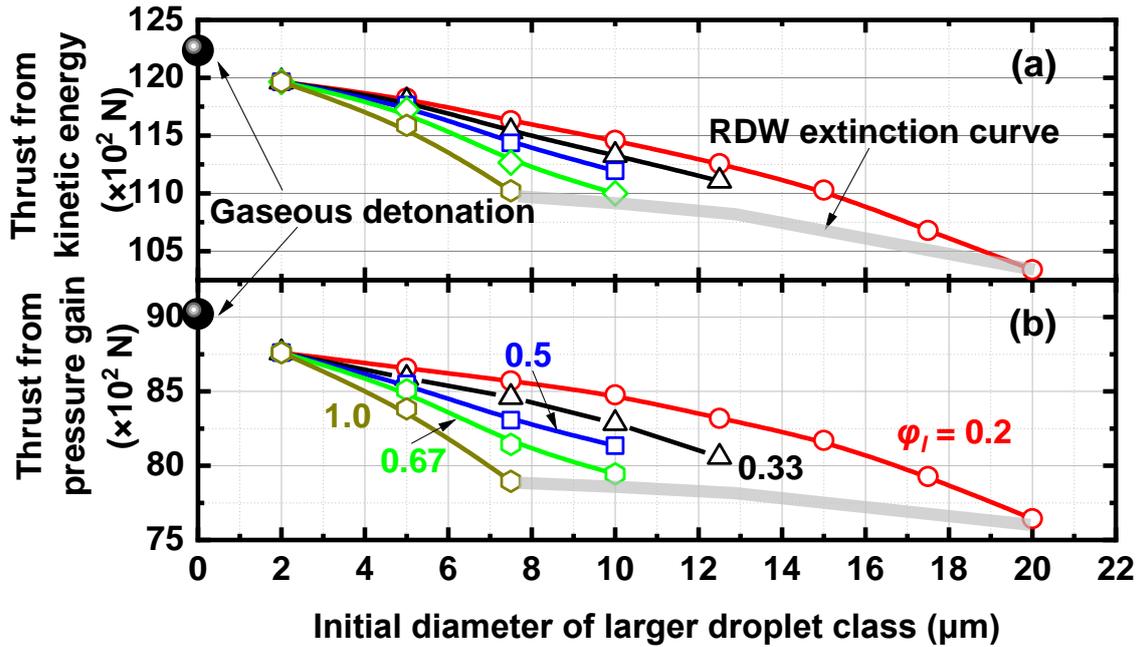

Fig. 17 Thrust force from (a) kinetic energy and (b) pressure gain. $d_s^0 = 2$ μm and $\phi_t = 1.0$.

## 4. Conclusions

Two-dimensional rotating detonations fueled by liquid *n*-heptane sprays are simulated with Eulerian – Lagrangian method. Bi-disperse fuel droplets without any fuel pre-vaporization are considered in our work and paramettric studies are performed to clarify the influences of liquid fuel droplet diameter and equivlance ratio on rotating detonation wave propagation, reactant mixing and propulsion performance.

In mono-sized sprays, when the droplet diameter is small (2 μm), the *n*-heptane droplets can completely evaporate in the fuel refilling area. While droplet diameter increases, a reflected shock can be obsevered after the detonation wave and the larger droplets can not completely evaporate in the fuel refilling area and exist behind the detonation wave. When the droplet diameter is larger than 10 μm, the higher pressure after the detonation wave leads to the reactants can not be sprayed into the combustor eventually leading to the extinction of the detonation wave. In bi-disperse sprays with 50% droplets of $d_s^0$ = 2 μm and 50% large droplets of $d_l^0$ = 10 μm, the presence of droplets with small diameter maintains the stable propagation of the detonation wave, while reflected shock is also observed.

The incomplete evaporation of droplets in the fuel filling area near the inlet of the RDC leads to an average equivalence ratio in the fuel refilling area ($\phi_{eff}$) lower than $\phi_t$. $\phi_{eff}$ decreases with increasing $d_l^0$ and the decreasing $\phi_{eff}$ leads to a decrease in the detonated fuel fraction with increased $d_l^0$. The detonation propagation speeds from liquid fueled RDC are lower than that of the corresponding gaseous RDC. The increase in $d_l^0$ and $\phi_l$ raise the velocity deficit. Finally, $d_l^0$ and $\phi_l$ also affect the thrust and specific impulse of the RDC. The propulsive performance decreases with increased $d_l^0$ and $\phi_l$.

## Acknowledgement

The simulations used the ASPIRE 1 Cluster from National Supercomputing Centre, Singapore (NSCC) (https://www.nscc.sg/). This work is supported by China Scholarship Council (No. 202006680045).